\documentstyle[12pt,epsfig]{article}

\def\gsim{\lower0.5ex\hbox{$\:\buildrel >\over\sim\:$}}
\def\lsim{\lower0.5ex\hbox{$\:\buildrel <\over\sim\:$}}
\def \n{\noindent}

\begin{document}
\baselineskip=18pt

\begin{center}
{\large\bf Extended Friedberg Lee
hidden symmetries, quark masses and CP-violation with four generations}
\end{center}

\vspace*{0.5 in}

\renewcommand{\thefootnote}{\fnsymbol{footnote}}
\centerline{Shaouly Bar-Shalom$^{a}$\footnote{Electronic address: shaouly@physics.technion.ac.il},
David Oaknin$^{a}$\footnote{Electronic address: d1306av@gmail.com},
Amarjit Soni$^b$\footnote{Electronic address: soni@bnl.gov}}

\vspace{0.1 in}

\centerline{\it $^a$Physics Department, Technion-Institute of Technology, Haifa 32000, Israel}
\centerline{\it $^b$Theory Group, Brookhaven National Laboratory, Upton, NY 11973, USA}

\renewcommand{\thefootnote}{\arabic{footnote}}
\setcounter{footnote}{0}
\vspace*{0.5 in}
\begin{abstract}
Motivated in part by the several observed anomalies involving CP asymmetries of B and
$B_s$ decays, we
consider the Standard Model with a 4th sequential family (SM4) which seems to 
offer a rather simple resolution.
We initially assume T-invariance by taking the up and down-quark
$4 \times 4$ mass matrix to be real. Following Friedberg and Lee (FL), we then
impose a ``hidden" symmetry on the unobserved (``hidden")
up and down-quark SU(2) states.
The hidden symmetry for four generations
ensures the existence of two zero-mass eigenstates,
which we take to be the $(u,c)$ and $(d,s)$
states in the up and down-quark sectors, respectively.
Then, we simultaneously break T-invariance and the hidden symmetry by
introducing two phase factors in each sector.
This breaking mechanism generates the small quark masses
$m_u,~m_c$ and
$m_d,~m_s$, which, along with the orientation of 
the hidden symmetry, determine the size of CP-violation in the SM4.
For
illustration we choose a specific physical picture for the hidden symmetry and 
the breaking mechanism that reproduces the observed
quark masses, mixing angles and CP-violation, and at the same 
time allows us to further obtain very interesting relations/predictions 
for the mixing angles of $t$ and $t^\prime$. For example,
with this choice we get 
$V_{td} \sim  (V_{cb}/V_{cd} - V_{ts}/V_{us}) +  {\cal O}(\lambda^2)$ and 
$V_{t^\prime b} \sim V_{t^\prime d} \cdot (V_{cb}/V_{cd})$,
$V_{t b^\prime} \sim V_{t^\prime d} \cdot (V_{ts}/V_{us})$, implying
that $V_{t^\prime d}  > V_{t^\prime b},V_{t b^\prime}$. We furthermore
find that the Cabibbo angle is related to the orientation of the hidden symmetry and 
that the key CP-violating quantity of our model
at high-energies, $J_{SM4} \equiv {\rm Im} \left( V_{tb} V_{t^\prime b}^\star V_{t^\prime b^\prime} V_{t b^\prime}^\star \right)$, which is the
high-energy analogue of the Jarlskog invariant of the SM, 
is proportional to the light-quark masses and the measured CKM angles:
$| J_{SM4} | \sim A^3 \lambda^5 \times ( \sqrt{ m_u/m_t }
+ \sqrt{ m_c/m_{t^\prime} }
- \sqrt{ m_d/m_b }
+ \sqrt{ m_s/m_{b^\prime}}) \sim
10^{-5}$, where $A \sim 0.81$ and $\lambda=0.2257$ are the Wolfenstein parameters. 
Other choices for the orientation of the hidden symmetry and/or the breaking mechanism 
may lead to different physical outcomes. A general solution, obtained numerically, 
will be presented in a forthcoming paper.
\end{abstract}

\newpage

\section{Introduction}

In spite of the success of the Standard Model (SM) in explaining almost all
of the observed phenomena in particle physics, it does not address some fundamental issues, such as the hierarchy problem, dark matter, the
matter anti-matter asymmetry in the universe etc. Also unexplained are the
issues in flavor physics, such as the hierarchy
of fermion masses and the number of families.
There are strong indications, from both the theoretical and experimental
points of views, that some of these unresolved questions are related to some new physics, maybe at the near by TeV-scale. It is, therefore, hoped that, with the LHC
turning on very soon, we will get a first hand glimpse of the new physics at the
TeV scale and new hints from nature to some of these issues and, in particular to the physics of flavor.

In this paper we wish to study some of the fundamental unresolved issues of flavor within a simple extension of the SM, in which a fourth sequential family of fermions is added - the SM4.
Indeed, the four generations scenario can play an important role in flavor physics
\cite{oldsoni}, and has recently gained some new interest as it might shed
 new light on baryogenesis and on CP-violation in K in B, $B_s$ decays \cite{hou2006,soni4gen,CPbaryo1,CPbaryo2,kribs}.
This model, which can be regarded as an effective low energy description of some higher energy and more fundamental underlying theory, retains all the features of the SM
with three generations (which from here on we will denote as SM3), except that it brings into existence the new heavy fermionic members
$t^\prime$ and $b^\prime$, which form the 4th quark doublet
and a similar leptonic doublet, where the ``neutrino" of the 4th family
must also be rather heavy, with mass $\gsim M_Z/2$. This may well be  an
important clue that the underlying nature of the 4th family may be quite
different from the 1st three families. This line of thinking may in fact lead to a
dark matter candidate \cite{0310006}.

The addition of the fourth generation to the SM3
means that the CKM matrix can now potentially have six independent real parameters/angles and three
physical CP-violating phases \cite{jarlskog}.
The two additional phases (with respect to the SM3) provide new sources of CP-violation and may, thus, give rise
to new CP-violating effects.
Indeed, in a recent paper \cite{soni4gen}, it was shown that
a fourth family of quarks with $m_{t^\prime}$ in the range of $\sim 400 - 600$ GeV provides a simple and perhaps rather natural explanation
for the several indications of new physics \cite{newsoni} that have been observed involving CP asymmetries in b-quark systems, and
this in fact forms an important motivation for our work.
Such heavy fermionic states
point to the interesting possibility that the 4th family may play a role in dynamical
electroweak symmetry breaking (EWSB), since the mechanism of dynamical mass generation seems to require 
such heavy masses \cite{dynEWSB,dynEWSB2}. In addition, as mentioned above, the
new CP-violating phases may play an important role in generating the
baryon asymmetry in the universe \cite{CPbaryo1,CPbaryo2}, which is difficult to address within
the SM3.

We note in passing that a 4th generation of quarks (and leptons) with such
heavy  masses
is not ruled out by precision electroweak constraints, but rather  requires
that correspondingly the Higgs has to be heavier, $\gsim$ 300 GeV \cite{kribs}.

In a recent paper \cite{FL} that also partly motivated the present work, Friedberg and Lee (FL) suggested
a very interesting new approach for the generation of CP-violation and quark masses in the SM3: that a weakly broken symmetry
which is operational in the SU(2) (weak)
fermionic states relates the smallness of CP-violation
to the smallness of the light-quark masses $m_d$ and $m_u$. More specifically, they imposed a ``hidden" symmetry
on the weak states of the quarks (named henceforward as the ``hidden" frame), which is then weakly broken by small CP-phases that generate
the non-zero masses for the light-quarks u and d. They found a very interesting
relation between CP-violation and the light-quark masses:
\begin{equation}
J_{SM} \propto \sqrt{\frac{m_d m_s}{m_b^2}} \label{FLrelation}~,
\end{equation}
\noindent where $J_{SM}$ is the Jarlskog invariant responsible for CP-violation in the SM3 \cite{jarlskog}.

The main appealing feature of the FL mechanism is that the CP-violating phases are the small parameters that control the breaking of the hidden symmetry and are, therefore, the generators of the small masses of the first generation quarks.
Unlike the conventional SM3 picture, the FL mechanism gives a physical meaning to the
 rotations of the quark fields (i.e., from the weak basis to the physical mass eigenstates basis) in the up and down quark sector separately, since there is an independent hidden symmetry for each sector.

As we will show in this paper, the idea of FL and their main result in Eq.~\ref{FLrelation} is extremely
interesting when applied
to the SM4 case and our extension will lead it to predictive power . In particular, with an appropriate choice of a hidden symmetry, it allows to generate {\it all four masses} of the $u,d,c$ and $s$-quarks in terms of the masses of the four heavy quarks $b,t,b^\prime$ and $t^\prime$ and the new CP-phases. It also gives distinct predictions for the 4th generation mixing angles and for the size of CP-violation in this theory, subject to the constraints coming from existing data on the SM3's $3 \times 3$ CKM matrix and quark masses. 
Thus, the hidden symmetry framework for the SM4 case can be directly tested
in collider experiments. In particular, we give distinct predictions for the new mixing angles and for the size of the new CP-violating quantities associated with the dynamics of the 4th generation quarks.

On the other hand, the construction of a hidden symmetry for the SM4 case, and
the generation of the four light-quark masses in conjunction with T-violation, is more challenging and rather intricate and analytically involved than in the case of the SM3. This is mainly due to the fact that the phase-space of the hidden symmetry in the SM4 case is much broader and that, as opposed to the FL mechanism for the SM3
where the CP-phases generate only the masses of the 1st generation fermions, here we use the new CP-phases (of the SM4) as generators of all four light-quark masses $m_d,m_u,m_s,m_c$, which makes it more difficult to find a physical solution. To put it in another way, our hidden symmetry for the SM4 case defines
a plane in which the theory is invariant whereas for three families the symmetry
is ``one-dimensional", i.e., defines a direction/vector.

In order to spell out our notation and the general formalism of the hidden symmetry and its
breaking mechanism within the SM4, we first consider the $4 \times 4$ up and down-quark
Yukawa terms in the SM4 (after EWSB):

\begin{eqnarray}
{\cal M}(q^{u,d}) =
\left(q^{u,d}_1,~q^{u,d}_2,~q^{u,d}_3,~q^{u,d}_4 \right) M(q^{u,d})
\left( \begin{array}{c}
q^{u,d}_1 \\ q^{u,d}_2 \\ q^{u,d}_3 \\ q^{u,d}_4 \end{array} \right) ~,
\end{eqnarray}

\n where $q^{u,d}_i$, $i = 1-4$, are the hidden SU(2) quark states of the SM4,
and $M(q^{u,d})$ are the corresponding mass matrices in the hidden frame basis.

As our zeroth-approximation we assume invariance under time reversal, thus taking $M_0(q^{u,d})$ (the subscript $0$ will henceforward denote the zeroth-order
quantities) to be real and symmetric. We can then extend FL's idea to the case of the SM4 by ``doubling" the hidden symmetry in each quark sector
(in the following we drop the indices $u$ and $d$, where
unless stated otherwise, it is understood that the discussion below applies to both up and down sectors):

\begin{eqnarray}
&&q_1 \to q_1 + \delta^1_z z + \delta^1_t t~, \nonumber \\
&&q_2 \to q_2 + \delta^2_z z + \delta^2_t t ~,  \nonumber \\
&&q_3 \to q_3 + \delta^3_z z + \delta^3_t t ~,  \nonumber \\
&&q_4 \to q_4 + \delta^4_z z + \delta^4_t t \label{HS} ~,
\end{eqnarray}

\n where $z$ and $t$ are space-time independent constants
of Grassmann algebra anticomuting with the Dirac field operators,
and $\delta^i_z,~\delta^i_t$ are c-numbers.

Since $M_0(q)$ is a real symmetric $4 \times 4$ matrix, 
it is  characterized in general by 10 real parameters. However,
imposing the hidden symmetry in Eq.~\ref{HS} eliminates 2 of the 10 parameters. The hidden symmetry of Eq.~\ref{HS} ensures (under the invariance of ${\cal M}_0(q^{u,d})$) the existence of two massless quark states in each sector, which we will identify as
$m_u$ and $m_c$ (in the up-quark sector) and as $m_d$ and $m_s$
(in the down-quark sector).
The corresponding two massless eigenvectors of $M_0(q)$ are thus identified as
the zeroth-order $u$ and $c$ states, $v_u^0$ and $v_c^0$ (with $m_u^0,~m_c^0=0$)
and in the down-quark sector as
the zeroth-order $d$ and $s$ states, $v_d^0$ and $v_s^0$ (with $m_d^0,~m_s^0=0$).
That is, since nature proves to have a large hierarchical mass structure in the quark sector,
we will consider the SM4 in the chiral limit for the first two generations of quarks -
$m_{u,d,c,s} =0$. Accordingly, the two massive eigenvectors are identified as
the zeroth-order $t$ and $t^\prime$ states (or $b$ and $b^\prime$ states) $v_t^0$ and $v_{t^\prime}^0$ , (or $v_b^0$ and $v_{b^\prime}^0$) with
masses (i.e., eigenvalues) $m^0_t,~m^0_{t^\prime}$ (or $m^0_b,~m^0_{b^\prime}$).
In particular, it is easy to show that in the hidden basis
$\{q_1,q_2,q_3,q_4\}$ the massless eigenvectors
span a 2-dimensional subspace of the form:

\begin{eqnarray}
v_u^0,v_c^0 \in \left( \begin{array}{c}
\delta^1_z \\ \delta^2_z \\ \delta^3_z \\ \delta^4_z \end{array} \right) ,
\left( \begin{array}{c}
\delta^1_t \\ \delta^2_t \\ \delta^3_t \\ \delta^4_t \end{array} \right) ~,
\end{eqnarray}

\noindent and similarly in the down-quark sector.

The next step towards establishing the complete physical picture of quark masses and mixings is to simultaneously break T-invariance and
the hidden symmetry by inserting two new phase factors into $M_0$, in each sector.
In the following we will construct a general framework that defines the hidden symmetry
in the SM4 scenario in a form that emphasizes the underlying geometrical picture,
and, then, give a concrete physical example for the breaking
mechanism.

\section{Hidden symmetry, T-invariance and the zeroth-order spectrum for the SM4}

In a generalization of the FL idea to the case of the SM4,
let us assume, at the first stage that the zeroth-order mass matrix $M_0$
is real and invariant under the following translational symmetry (we will denote
this symmetry as Hidden Symmetry 1, HS1)
\begin{eqnarray}
\nonumber
q_1 & \rightarrow & q_1 + c_\theta z ~,\nonumber \\
q_2 & \rightarrow & q_2 + s_\theta c_\phi z ~,\nonumber \\
q_3 & \rightarrow & q_3 + s_\theta s_\phi c_\omega z ~, \nonumber \\
q_4 & \rightarrow & q_4 + s_\theta s_\phi s_\omega z \label{HS1} ~.
\end{eqnarray}
where $c_\theta,s_\theta = \cos\theta, \sin\theta$ etc., and
$z$ is a space-time independent constant
of Grassmann algebra anticomuting with the Dirac fields.

This symmetry guarantees that the vector
\begin{eqnarray}
Q_1 = c_\theta q_1 + s_\theta c_\phi q_2 +  s_\theta s_\phi c_\omega q_3 + s_\theta s_\phi s_\omega q_4 \label{Q0} ~,
\end{eqnarray}
is a massless eigenstate of the theory, as under the HS1 it transforms as
$Q_1 \rightarrow Q_1 + z$.
On the other hand, the three orthogonal (to $Q_1$) vectors
\begin{eqnarray}
Q_2 &=& -s_\theta q_1 + c_\theta c_\phi q_2 + c_\theta s_\phi c_\omega q_3 + c_\theta s_\phi s_\omega q_4 \nonumber \\
Q_3 &=& -s_\phi q_2 + c_\phi c_\omega q_3 + c_\phi s_\omega q_4 \nonumber \\
Q_4 &=& -s_\omega q_3 + c_\omega q_4 \label{Q123}~,
\end{eqnarray}
are invariant under the HS1, i.e., $Q_i \to Q_i$ for $i=2,3,4$.
The rotation from the hidden frame $\{q_1,q_2,q_3,q_4\}$ to
the HS1 frame $\{Q_1,Q_2,Q_3,Q_4\}$ can be written as $Q_i = R_{ij} q_j$, thus defining the real
unitary matrix $R$:

\begin{eqnarray}
R =\left( \begin{array}{cccc}
c_\theta & s_\theta c_\phi &  s_\theta s_\phi c_\omega & s_\theta s_\phi s_\omega \\
-s_\theta & c_\theta c_\phi & c_\theta s_\phi c_\omega & c_\theta s_\phi s_\omega \\
0 & -s_\phi & c_\phi c_\omega & c_\phi s_\omega \\
0 & 0 & -s_\omega & c_\omega
\end{array} \right) \label{Rmatrix}~.
\end{eqnarray}

Demanding translational invariance under HS1 of Eq.~\ref{HS1},
$M_0$ has only one massless eigenstate (the state $Q_1$).
Thus, in order to enforce the chiral limit for the first two generations,
we will demand that the zeroth-order mass matrix is invariant under an additional translation operation, which is operational in the
HS1 frame $\{Q_1,Q_2,Q_3,Q_4\}$ and which
we will name Hidden Symmetry 2 (HS2). Without loss of generality, we assume that HS2 is orthogonal to HS1 as follows:

\begin{eqnarray}
\nonumber
Q_1 & \rightarrow & Q_1 ~, \nonumber \\
Q_2 & \rightarrow & Q_2 + c_\zeta  t ~, \nonumber\\
Q_3 & \rightarrow & Q_3 + s_\zeta  c_\eta t ~, \nonumber \\
Q_4 & \rightarrow & Q_4 + s_\zeta s_\eta t \label{HS2} ~.
\end{eqnarray}

The additional symmetry HS2 guarantees that the vector
\begin{equation}
P_1 = c_\zeta Q_2 + s_\zeta c_\eta Q_3 + s_\zeta s_\eta Q_4
\label{P0} ~,
\end{equation}
which is orthogonal to $Q_1$, is also massless.

The most general form of the Yukawa term ${\cal M}_{0}$ that is invariant under the independent translations in both directions HS1 and HS2, can then be written as:
\begin{eqnarray}
{\cal M}_{0} = \alpha | c_\eta Q_4 - s_\eta Q_3 |^2  +
\beta | c_\zeta Q_4 - s_\zeta s_\eta Q_2 |^2 + \gamma | c_\zeta Q_3 - s_\zeta c_\eta Q_2|^2
\label{zeroth_order_matrx}~,
\end{eqnarray}
and this defines the quark mass matrix $M_0$. Recall that, since $M_0$ is invariant under HS1 and HS2, two of its four eigenstates, i.e., $Q_1$ and $P_1$, are necessarily
massless.

Before deriving the full zeroth-order system (i.e., 2 non-zero masses and 4 states), we wish to point out
the mapping of our double hidden symmetry (HS1 and HS2) to the generic parameterizations of the hidden symmetry in Eq.~\ref{HS}. In particular, using the definition for HS1 and HS2 in
Eqs.~\ref{HS1} and \ref{HS2}, respectively, and the fact that $q = R^{-1}Q$,
we obtain the overall hidden symmetry for
the SM4 case:
\begin{eqnarray}
\nonumber
q_1 & \rightarrow & q_1 + c_\theta z - s_\theta c_\zeta t ~, \nonumber \\
q_2 & \rightarrow & q_2 + s_\theta c_\phi z + \left[c_\theta c_\phi c_\zeta - s_\phi s_\zeta c_\eta \right]  t ~, \nonumber \\
q_3 & \rightarrow & q_3 + s_\theta s_\phi c_\omega z +
\left[c_\theta  s_\phi c_\omega c_\zeta + c_\phi c_\omega s_\zeta c_\eta - s_\omega s_\zeta s_\eta \right] t ~, \nonumber \\
q_4 & \rightarrow & q_4 + s_\theta s_\phi s_\omega z +
\left[c_\theta s_\phi s_\omega c_\zeta + c_\phi s_\omega s_\zeta c_\eta + c_\omega s_\zeta s_\eta \right] t \label{fullHS} ~,
\end{eqnarray}
from which one can extract the hidden symmetry
parameters $\delta_z^i$ and $\delta_t^i$ of Eq.~\ref{HS}, as a function
of the angles which define the orientations of HS1 and HS2 with respect to the hidden frame $\{ q_1,q_2,q_3,q_4 \}$.

Note that the expression for ${\cal M}_{0}$ in Eq.~\ref{zeroth_order_matrx}
contains five angles:
the two (explicit) angles $\zeta,\eta$ associated with the orientation of HS2 with respect to the HS1 frame $\{Q_1, Q_2, Q_3, Q_4 \}$ and the three angles $\theta,\phi,\omega$ associated with
the orientation of HS1 with respect to the hidden frame $\{q_1, q_2, q_3, q_4 \}$, which enter through the rotation $Q=Rq$.
Thus, along with the parameters $\alpha,\beta$ and $\gamma$,
${\cal M}_{0}$ in Eq.~\ref{zeroth_order_matrx} is parameterized by 8 real parameters
(in each sector) as required when imposing the double hidden symmetry
(see discussion above).
However, there is one non-physical angle in {\it each sector} which results from
the fact that the two orthogonal states $Q_1,P_1$ are massless at
zeroth-order and are, therefore, indistinguishable.
This can be easily understood by considering
the geometrical interpretation of the hidden symmetry in the SM4
case. In particular, the double hidden symmetry
(HS1+HS2) defines a plane in the hidden frame $\{q_1, q_2, q_3, q_4 \}$ under which the theory is invariant.
This is the plane spanned by the two orthogonal vectors $Q_1$ and $P_1$.
We, therefore, have the freedom to make any unitary transformation in
the $Q_1-P_1$ plane/subspace (in both up and down-quark sectors)
without affecting the physical picture. This allows us to
eliminate one angle
in each of the $(v_d^0,v_s^0)$ and $(v_u^0,v_c^0)$ subspaces.
Thus, without loss of generality we find it convenient
to choose $\omega=\pi/2$ in both sectors,
which sets $Q_4=q_3$ and $Q_1,Q_2,Q_3 \perp q_3$. This is analogous to a gauge condition
in a vector field theory as also identified in \cite{FL}.
Note that even though at each sector the massless states $(v_d^0,v_s^0)$ and $(v_u^0,v_c^0)$ are
indistinguishable at the zeroth-order, as we will see in the next section,
after breaking the hidden symmetry this degeneracy is removed,
and those (now massive) states become well defined.

We are now ready to derive the mass spectrum and the
$4 \times 4$ CKM matrix at zeroth-order, i.e., without T-violation. Recall that, by construction, there are two massless states, given by $Q_1$ and $P_1$.
In order to find the 2 massive states we can apply the original FL formulae for three generations to
the $\{ Q_2, Q_3, Q_4 \}$ subspace.
As in \cite{FL}, we find that the eigensystem of $M_{0}$ depends
only on two linear combinations of $\alpha,\beta,\gamma$, so that one
of these three parameters can be ``gauged away". Following the choice of FL in \cite{FL}, we eliminate the parameter $\gamma$ using the ``gauge" condition (i.e., this has no effect on the physical outcome):
\begin{eqnarray}
\frac{\beta}{\gamma} = 1 \label{gauge1}~.
\end{eqnarray}

Using this condition, we diagonalize the mass matrix
$M_0$ and find that the two massive states are:
\begin{eqnarray}
P_2 &=& -s_\zeta Q_2 + c_\zeta c_\eta Q_3 + c_\zeta s_\eta Q_4 ~, \nonumber \\
P_3 &=& -s_\eta Q_3 + c_\eta Q_4 \label{P23}~,
\end{eqnarray}
with masses:
\begin{eqnarray}
m_{P_2}&=& \beta ~, \\
m_{P_3}&=&\alpha  + c_\zeta^2 \beta =
\alpha + c_\zeta^2 m_{P_2}   \label{M23}~.
\end{eqnarray}

Note that, for $m_{P_3} >> m_{P_2}$ and/or $c_\zeta \to 0$ we have
$m_{P_3} \approx \alpha$ and $m_{P_2} \approx \beta$ (see below).

Thus the complete set of eigenstates of $M_0$ at zeroth-order becomes quite simple, as it is given by $\{ Q_1, P_1, P_2, P_3 \}$ with masses $\{ 0, 0, m_{P_2}, m_{P_3} \}$, which we hanceforward identify (in each sector) as the zeroth-order quark states:
\begin{eqnarray}
\{ v_d^0, v_s^0, v_b^0, v_{b^\prime}^0 \}
&\equiv& \{ Q_1^d, P_1^d, P_2^d, P_3^d \} \label{rel1}~, \\
\{ v_u^0, v_c^0, v_t^0, v_{t^\prime}^0 \}
&\equiv& \{ Q_1^u, P_1^u, P_2^u, P_3^u \} \label{rel2}~,
\end{eqnarray}
with masses $m_d^0=m_s^0=m_u^0=m_c^0=0$ and:
\begin{eqnarray}
&&m_b^0 =\beta_d,~ m_{b^\prime}^0  \approx \alpha_d ~, \nonumber \\
&&m_t^0 =\beta_u,~ m_{t^\prime}^0  =
\alpha_u + c_{\zeta_u}^2  m_t^0 \label{zeromasses}~,
\end{eqnarray}
where the superscripts $d$ and $u$ distinguish between the parameters
in the down-quark and up-quark sectors, respectively.
Note that since T-violation is responsible
for generating the light-quark masses, it is a small perturbation to the T-invariant zeroth-order spectrum. Thus for all practical purposes we can set: $m_b \approx m_b^0$,
$m_{b^\prime} \approx m_{b^\prime}^0$,
$m_t \approx m_t^0$ and $m_{t^\prime} \approx m_{t^\prime}^0$ (see also below).

Using the orientation of the HS1 frame $Q_i$ with respect to the hidden frame $q_i$, i.e., $Q=Rq$ with $R$ given in Eq.~\ref{Rmatrix},
and the orientation
of the states $P_{2},P_{3}$ with respect to the $\{ Q_2, Q_3, Q_4 \}$ subframe (as given in Eq.~\ref{P23}),
we can write the set of four eigenstates in each sector
in terms of the weak (hidden) states $q_i$ (as required in order to derive the zeroth-order (real) $4\times 4$ CKM matrix):
\begin{eqnarray}
\left( \begin{array}{c}
v_d^0 \\ v_s^0 \\ v_b^0 \\ v_{b^\prime}^0 \end{array} \right) =
\left( \begin{array}{c}
R^d_{1i} \\ A^d_i  \\ B^d_i  \\ C^d_i  \end{array} \right) q^d_i ~,~
\left( \begin{array}{c}
v_u^0 \\ v_c^0 \\ v_t^0 \\ v_{t^\prime}^0 \end{array} \right) =
\left( \begin{array}{c}
R^u_{1i} \\ A^u_i  \\ B^u_i  \\ C^u_i  \end{array} \right) q^u_i \label{rel3} ~,
\end{eqnarray}
where the superscripts $u$ and $d$ are again added in order to distinguish between the angles associated with the up and down-quark sectors, respectively.
Also,
\begin{eqnarray}
A^d_i &\equiv& \cos\zeta_d \cdot R^d_{2i} + \sin\zeta_d\cdot \cos\eta_d\cdot R^d_{3i} + \sin\zeta_d\cdot \sin\eta_d\cdot R^d_{4i} \label{ai} ~ \\
B^d_i &\equiv& -\sin\zeta_d \cdot R^d_{2i} + \cos\zeta_d\cdot\cos\eta_d\cdot R^d_{3i} + \cos\zeta_d\cdot \sin\eta_d\cdot R^d_{4i} \label{bi} ~, \\
C^d_i &\equiv& -\sin\eta_d\cdot R^d_{3i} + \cos\eta_d\cdot R^d_{4i} \label{ci}~,
\end{eqnarray}
and similarly for $A^u_i,B^u_i,C^u_i$ using $R^u$ and $\zeta_u,\eta_u$.

Then denoting by $D_0 = (v_d^0, v_s^0, v_b^0, v_{b^\prime}^0 )$ and
$U_0 = (v_u^0, v_c^0, v_t^0, v_{t^\prime}^0)$
the unitary matrices that diagonalize the real and symmetric mass matrices
in the down and up-quark sectors, respectively:
\begin{eqnarray}
D_0^\dagger M_0(q^d) D_0 &=& {\rm diag}(0,0,m_b^0,m_{b^\prime}^0) ~, \\
U_0^\dagger M_0(q^u) U_0 &=& {\rm diag}(0,0,m_t^0,m_{t^\prime}^0) ~,
\end{eqnarray}
\noindent we can obtain the $4\times 4$ zeroth-order CKM matrix of the SM4
(i.e., without T-violation):
\begin{eqnarray}
V^0(CKM) = U_0^\dagger D_0 \label{CKM0}~.
\end{eqnarray}

The general expression for $V^0(CKM)$ in terms of the angles that define
the hidden symmetry in the up and down-quark sectors is rather complicated
to be written here.
Let us, therefore, choose a specific physical
orientation of the hidden symmetry, where the direction of HS2 is partly fixed by the angle $\zeta$ with the choice $\zeta =\omega=\pi/2$ in each sector (recall that we have fixed the angle $\omega = \pi/2$ in a manner similar to choosing a gauge).
This orientation is physically viable in the sense that it reproduces 
the observed light-quark masses and the measured CKM mixing angles. It will be used in the next sections
to demonstrate the general mechanism for breaking the hidden symmetry and T-invariance and the corresponding generation of the light-quark masses.

In particular, using Eqs.~\ref{rel1}-\ref{CKM0} with $\zeta =\omega=\pi/2$ we obtain:
\begin{eqnarray}
V_{ud}^0 &=& c_{\theta_u} c_{\theta_d} + s_{\theta_u} s_{\theta_d} \cos(\phi_u-\phi_d) ~,\nonumber\\
V_{us}^0 &=& s_{\theta_u} c_{\eta_d} \sin(\phi_u-\phi_d) ~, \nonumber \\
V_{ub}^0 &=& c_{\theta_u} s_{\theta_d} - s_{\theta_u} c_{\theta_d} \cos(\phi_u-\phi_d) ~,\nonumber\\
V_{ub^\prime}^0 &=& - s_{\theta_u} s_{\eta_d} \sin(\phi_u-\phi_d) ~,\nonumber\\
V_{cd}^0&=&- s_{\theta_d} c_{\eta_u} \sin(\phi_u-\phi_d) ~,\nonumber\\
V_{cs}^0 &=& s_{\eta_u} s_{\eta_d} + c_{\eta_u} c_{\eta_d} \cos(\phi_u-\phi_d) ~,\nonumber\\
V_{cb}^0&=& c_{\eta_u} c_{\theta_d} \sin(\phi_u-\phi_d) ~,\nonumber\\
V_{cb^\prime}^0 &=& s_{\eta_u} c_{\eta_d} - c_{\eta_u} s_{\eta_d} \cos(\phi_u-\phi_d) ~,\nonumber\\
V_{td}^0&=& c_{\theta_d} s_{\theta_u} - s_{\theta_d} c_{\theta_u} \cos(\phi_u-\phi_d) ~,\nonumber\\
V_{ts}^0&=& -c_{\eta_d} c_{\theta_u} \sin(\phi_u-\phi_d) ~,\nonumber\\
V_{tb}^0 &=& s_{\theta_u} s_{\theta_d} + c_{\theta_u} c_{\theta_d} \cos(\phi_u-\phi_d) ~,\nonumber\\
V_{tb^\prime}^0&=& s_{\eta_d} c_{\theta_u} \sin(\phi_u-\phi_d) ~,\nonumber\\
V_{t^\prime d}^0&=& s_{\theta_d} s_{\eta_u} \sin(\phi_u-\phi_d) ~,\nonumber\\
V_{t^\prime s}^0&=&s_{\eta_d} c_{\eta_u} - c_{\eta_d} s_{\eta_u} \cos(\phi_u-\phi_d) ~,\nonumber\\
V_{t^\prime b}^0&=& - s_{\eta_u} c_{\theta_d} \sin(\phi_u-\phi_d) ~,\nonumber\\
V_{t^\prime b^\prime}^0 &=& c_{\eta_d} c_{\eta_u} + s_{\eta_d} s_{\eta_u} \cos(\phi_u-\phi_d) \label{CKMel}~.
\end{eqnarray}

From these expressions we can find the size of some of the hidden symmetry angles in terms of the observed $3 \times 3$ CKM elements and, also, several interesting and
surprising relations/predictions for the mixing angles of the 4th generation quarks with the first 3 generations:
\begin{eqnarray}
- \tan\theta_u &=& \frac{V_{us}}{V_{ts}} = \frac{V_{ub^\prime}}{V_{tb^\prime}} \label{eqV1}~,\\
- \tan\theta_d &=& \frac{V_{cd}}{V_{cb}} = \frac{V_{t^\prime d}}{V_{t^\prime b}} \label{eqV2}~,\\
- \tan\eta_u &=& \frac{V_{t^\prime d}}{V_{cd}} ~,\\
- \tan\eta_d &=& \frac{V_{u b^\prime}}{V_{us}} \label{eqV4} ~,
\end{eqnarray}
implying $V_{t^\prime d} > V_{t^\prime b}$ and
$V_{ub^\prime} > V_{tb^\prime}$ - opposite to the hierarchical
pattern as observed in the SM3's $3 \times 3$ block.

In addition,
taking $V_{ts}^2/V_{us}^2 \sim V_{cb}^2/V_{cd}^2  \ll 1$,
$V_{ud} \sim 1 - \lambda^2/2$ and $V_{cs} \sim 1 - \lambda^2/2$,
where $\lambda \sim 0.2257$ is the Wolfenstein parameter \cite{PDG},
we find that $\phi_u-\phi_d$ is the Cabibbo angle (i.e., the Wolfenstein parameter) with:
\begin{eqnarray}
\sin(\phi_u - \phi_d) &\sim& \lambda \sim 0.2257 ~, \\
\cos(\phi_u - \phi_d) &\sim& V_{ud} - {\cal O}(\lambda^2) ~,
\end{eqnarray}
and
\begin{eqnarray}
c_{\theta_d} &\sim& \frac{V_{cb}}{V_{cd}} \sim {\cal O}(\lambda) \label{ctetd}~\\
c_{\theta_u} &\sim& \frac{V_{ts}}{V_{us}} \sim {\cal O}(\lambda) \label{ctetu}~\\
\cos(\eta_u - \eta_d) &\sim& V_{cs} - {\cal O}(\lambda^2) \label{ceta}~,
\end{eqnarray}
also implying that $\eta_u \sim \eta_d$. This in turn gives:
\begin{eqnarray}
V_{t^\prime b^\prime} &\sim& V_{cs} ~\\
V_{u b^\prime} &\sim& V_{t^\prime d} \label{vubprime}~.
\end{eqnarray}

Furthermore, for the top-quark mixing angles we get:
\begin{eqnarray}
V_{tb} &\sim& 1 - {\cal O}(\lambda^2) ~,\\
V_{td} &\sim& \left( \frac{V_{cb}}{V_{cd}} - \frac{V_{ts}}{V_{us}} \right) + {\cal O}(\lambda^2) \label{eqV6} ~,
%V_{ts} &\sim& - \lambda c_{\theta_u} \sqrt{ 1 + 2 V_{ud}
%\frac{c_{\theta_d} c_{\theta_u}}{\lambda^2} } ~.
\end{eqnarray}

In the next sections we will use this physical setup to break T-invariance
and derive the CP-violating parameters of the model.

\section{T-violation and hidden symmetry breaking mechanism}

There are, of course, several ways to break the hidden symmetry
without breaking T-invariance.
Here we wish to extend the attractive mechanism for the simultaneous breaking
of both the hidden symmetry and T-invariance, that was suggested by Friedberg and Lee in \cite{FL} in the SM3 case, and formulate the general breaking mechanism for the SM4 case.

In particular, when the hidden symmetry and T-invariance are broken simultaneously,
the massless states $v_d^0,v_s^0,v_u^0,v_c^0$ (which were protected by the hidden symmetry) acquire a mass which is directly related to the size of the phases responsible for T-violation: two CP-violating phases in the up-quark
sector are needed to generate the masses $m_u$ and $m_c$, while two CP-violating phases in the down-quark sector generate the masses $m_d$ and $m_s$.
Since we know that
$m_{u,c} << m_{t,t^\prime}$ and $m_{d,s} << m_{b,b^\prime}$, we can treat the effect of
T-violation as a perturbation to the zeroth-order (T-invariant) approximation
in both the down and up-quark sectors.

In what follows we will describe the breaking mechanism using the generic notation outlined in the previous section, which holds for both down and up-quark sectors. The application of the results below to a specific sector is straight
forward.

In order to break the hidden symmetry we rewrite the zeroth-order Yukawa term ${\cal M}_0$ in terms of its eigenstates:

\begin{eqnarray}
{\cal M}_0 = \sum_i m_i^0 |v_i^0|^2 = m_{P_2} |P_2|^2 + m_{P_3} |P_3|^2~,
\end{eqnarray}
where we have used the fact that $m_{Q_1}=m_{P_1}=0$. This gives (see
Eqs.~\ref{rel1},\ref{rel2} and \ref{rel3}):
\begin{eqnarray}
(M_0)_{ij} = m_{P_2} B_i B_j + m_{P_3} C_i C_j \label{M0bc}~,
\end{eqnarray}
where we have dropped the superscripts $d$ or $u$ in the coefficients $B^{d,u}_i$ and $C^{d,u}_i$ (as defined in Eqs.~\ref{bi} and \ref{ci}), so that the expression above applies to both down and up sectors. T-invariance and the hidden symmetry
can then be broken by inserting a phase in any one of the non-diagonal entries
of $(M_0)_{ij}$ as follows:
\begin{eqnarray}
(\Delta M)_{ij} = \left( m_{P_2} B_i B_j + m_{P_3} C_i C_j \right) \cdot
\left( e^{i \delta_{ij}} - 1 \right) ~, j > i ~;~
(\Delta M)_{ji} = (\Delta M)_{ij}^\star
\label{deltaM}~,
\end{eqnarray}
such that
\begin{eqnarray}
M = M_0 + \Delta M ~.
\end{eqnarray}

We assume that $\delta_{ij} \ll 1$, hence, $\Delta M \ll M_0$ so that $\Delta M$ can be treated as a perturbation. As we shall
demonstrate in the next section,
in the minimal setup, two such phase
insertions (in each sector) are required in different locations in $M_0$ in order to break both HS1 and HS2
and to generate the observable masses
of the first 2 light generations of quarks. Thus, we can write
the overall T-violating term as:
\begin{eqnarray}
\Delta M \equiv \Delta M_z + \Delta M_t ~,
\end{eqnarray}
where $\Delta M_z$ and $\Delta M_t$
contain the new phases that break
HS1 and HS2, respectively, each given by the generic form in Eq.~\ref{deltaM}.
The T-violating mass term $\Delta M$ then shifts
the zeroth-order masses and states.
Using perturbation theory, these shifts are given in the general case without degeneracies  
by:

\begin{eqnarray}
\Delta m_q &\equiv& m_q - m_q^0 = (v_q^0)^\dagger \Delta M v_q^0 \label{dmass}~,
\end{eqnarray}
\begin{eqnarray}
\Delta v_q &\equiv& v_q- v_q^0 = \sum_{q\neq q^\prime} \frac{(v_{q^\prime}^0)^\dagger
\Delta M v_q^0}{m_q^0-m_{q^\prime}^0} v_{q^\prime}^0 \label{dvec}~,
\end{eqnarray}

\noindent where $m_q^0$ and $v_q^0$ are the zeroth-order masses and states
(i.e., $v_q^0$ and $v_{q^\prime}^0$ stands for any one of the vectors $Q_1,P_1,P_2,P_3$ in either the up or down-sectors), $\Delta m_q$  are the mass shifts due to the breaking of the hidden symmetry
and
$\Delta v_q$ contains the imaginary terms which are
$\propto i \sin\delta_{ij}$ from which the physical T-violating
elements of the $4 \times 4$ CKM matrix are constructed.
 
In our case, however, the states $Q_1$ and $P_1$ are degenerate. 
Thus, in order to find the physical masses ($m_{\pm}$) and their corresponding physical states ($v_{\pm}$) 
in the $Q_1 - P_1$ subspace, we need to diagonalize the
following $2 \times 2$ perturbation mass matrix in the $Q_1-P_1$ subspace:
\begin{eqnarray}
\Delta m(Q_1,P_1) =\left( \begin{array}{cc}
Q_1^\dagger \Delta M Q_1 & Q_1^\dagger \Delta M P_1  \\
P_1^\dagger \Delta M Q_1 & P_1^\dagger \Delta M P_1  \end{array} \right) \equiv
\left( \begin{array}{cc}
\Delta m_{QQ} & \Delta m_{QP}   \\
\Delta m_{PQ}  & \Delta m_{PP}   \end{array} \right)
\label{deltaQP}~,
\end{eqnarray}
where $ \Delta m_{QP} = (\Delta m_{PQ})^\dagger$ and $\Delta m_{QQ},~\Delta m_{PP}$ are real.     
That is, after breaking T-invariance,
the physical masses and states of the first two generations
are given by:
\begin{eqnarray}
m_\pm &=& \frac{\Delta m_{QQ} +\Delta m_{PP}}{2}
\left[ 1 \pm \sqrt{1 -  \frac{4 \left( \Delta m_{QQ} \Delta m_{PP} -
\Delta m_{QP} \Delta m_{PQ} \right)}{\left( \Delta m_{QQ} +\Delta m_{PP} \right)^2}} \right] \label{mpm}~,
\end{eqnarray}
and
\begin{eqnarray}
v_+ &=& \frac{1}{\sqrt{ \left| \Delta m_{QP} \right|^2 + \left(m_+ - \Delta m_{QQ} \right)^2}}
\left[ \Delta m_{QP} Q_1 + \left(m_+ - \Delta m_{QQ} \right) P_1  \right] \nonumber ~\\
v_- &=& \frac{1}{\sqrt{ \left|\Delta m_{PQ} \right|^2 + \left(m_- - \Delta m_{PP} \right)^2}}
\left[ \left(m_- - \Delta m_{PP} \right) Q_1 + \Delta m_{PQ} P_1  \right] \label{vpm}~.
\end{eqnarray}

The corresponding corrections/shifts to the physical states are still 
calculated from Eq.~\ref{dvec}, where now $v_q^0,v_{q^\prime}^0 \in \{v_-,v_+,P_2,P_3\}$. 
In particular, 
let us further define the ``perturbation matrix":
\begin{eqnarray}
(v_q^0)^\dagger \Delta M v_{q^\prime}^0 \equiv i P_{q q^\prime}
\label{pijdef}~,
\end{eqnarray}
where
$q \neq q^\prime$ and, to ${\cal O}(\delta)$, $P_{q q^\prime}$ are real and $P_{q^\prime q} = - P_{q q^\prime}$. That is,
$(v_{q^\prime}^0)^\dagger \Delta M v_q^0 = \left[(v_q^0)^\dagger \Delta M v_{q^\prime}^0 \right]^\dagger = -i P_{q q^\prime}$,
where $q,q^\prime \in d,s,b,b^\prime$ in the down-quark sector and $q,q^\prime  \in u,c,t,t^\prime$ in the up-quark sector. Also note that the perturbation matrix is diagonal in the $(v_- - v_+)$ subspace to ${\cal O}(\delta)$
(i.e., $P_{d s} = P_{sd} \approx {\cal O}(\delta^2)$ and $P_{uc} = P_{cu} \approx {\cal O}(\delta^2)$).

In the next section, for simplicity we will consider the case where the perturbation is diagonal in the
$Q_1-P_1$ subspace, i.e., $\Delta m_{QP} =0$ in Eq.~\ref{deltaQP}, so that $v_-=Q_1$ and 
$v_+ = P_1$. In this simple case we can use
Eqs.~\ref{dvec} and \ref{pijdef} to obtain the
${\cal O}(\delta)$ shifts, $\Delta v_q$, to the zeroth-order states
$(v_d^0, v_s^0, v_b^0, v_{b^\prime}^0 )$ and
$(v_u^0, v_c^0, v_t^0, v_{t^\prime}^0)$ (as defined in Eqs.~\ref{rel3}-\ref{ci}):
\begin{eqnarray}
\begin{array}{cc}
\Delta v_d = i \left( \frac{P_{d b}}{m_b} v_b^0 +
\frac{P_{d b^\prime}}{m_{b^\prime}} v_{b^\prime}^0 \right) &
\Delta v_u = i \left( \frac{P_{u t}}{m_t} v_t^0 +
\frac{P_{u t^\prime}}{m_{t^\prime}} v_{t^\prime}^0 \right) \\
\Delta v_s = i \left( \frac{P_{s b}}{m_b} v_b^0 +
\frac{P_{s b^\prime}}{m_{b^\prime}} v_{b^\prime}^0 \right) &
\Delta v_c = i \left( \frac{P_{c t}}{m_t} v_t^0 +
\frac{P_{c t^\prime}}{m_{t^\prime}} v_{t^\prime}^0 \right) \\
\Delta v_b = i \left( \frac{P_{d b}}{m_b} v_d^0 +
\frac{P_{s b}}{m_b} v_s^0 +
\frac{P_{b b^\prime}}{m_{b^\prime} - m_b} v_{b^\prime}^0 \right) &
\Delta v_t = i \left( \frac{P_{u t}}{m_t} v_u^0 +
\frac{P_{c t}}{m_t} v_c^0 +
\frac{P_{t t^\prime}}{m_{t^\prime} - m_t} v_{t^\prime}^0 \right) \\
\Delta v_{b^\prime} = i \left( \frac{P_{d b^\prime}}{m_{b^\prime}} v_d^0 +
\frac{P_{s b^\prime}}{m_{b^\prime}} v_s^0 +
\frac{P_{b b^\prime}}{m_{b^\prime} - m_b} v_{b}^0 \right) &
\Delta v_{t^\prime} = i \left( \frac{P_{u t^\prime}}{m_{t^\prime}} v_u^0 +
\frac{P_{c t^\prime}}{m_{t^\prime}} v_c^0 +
\frac{P_{t t^\prime}}{m_{t^\prime} - m_t} v_{t}^0 \right)
\label{deltav}
\end{array}
\end{eqnarray}
such that, to ${\cal O}(\delta)$, the physical states 
are given by $v_q = v_q^0 + \Delta v_q$. The corresponding 
${\cal O}(\delta)$ corrections to $v_q^0$ in the general case where the perturbation is
not diagonal in the $Q_1-P_1$ subspace, can be easily derived from 
the expressions
for $v_\pm$ in Eq.~\ref{vpm} and the shifts $\Delta v_q$ in Eq.~\ref{deltav} above. For example,   
\begin{eqnarray}
\Delta v_- &=& \frac{1}{\sqrt{ \left|\Delta m_{PQ} \right|^2 + \left(m_- - \Delta m_{PP} \right)^2}}
\left[ \left(m_- - \Delta m_{PP} \right) \cdot \Delta v_d + \Delta m_{PQ} \cdot \Delta v_s  \right]
\end{eqnarray}
where $\Delta v_{d,s}$ are given in Eq.~\ref{deltav}.

The physical (T-violating) $4 \times 4$ CKM matrix elements are, therefore, given
symbolically by ($u$ and $d$ stand for any of the up and down-quark states, respectively):
\begin{eqnarray}
V_{ud} = (v_u)^\dagger \cdot v_d = V_{ud}^0 + (\Delta v_u)^\dagger \cdot v_d^0 +
v_u^0 \cdot \Delta v_d ~,
\end{eqnarray}
where $V_{ud}^0 = (v_u^0)^T \cdot v_d^0$ is the zeroth-order CKM matrix elements
and the terms $[ (\Delta v_u)^\dagger \cdot v_d^0 ],~[v_u^0 \cdot \Delta v_d]$, which are also
functions of the zeroth-order CKM elements, are readily obtained from
Eq.~\ref{deltav} above. For example, in the simple case where 
$v_-=Q_1$ and $v_+ = P_1$, $V_{ud}$ (i.e., now the (11) elements
of $V$) is given by
\begin{eqnarray}
V_{ud} = V_{ud}^0 + i \left[
\frac{P_{d b}}{m_b} V_{u b}^0 +
\frac{P_{d b^\prime}}{m_{b^\prime}} V_{u b^\prime}^0
-
\frac{P_{u t}}{m_t} V_{td}^0 -
\frac{P_{u t^\prime}}{m_{t^\prime}} V_{t^\prime d}^0
\right] + {\cal O}(\delta^2) \label{Vudfull}~.
\end{eqnarray}

Note that the zeroth-order elements $V_{ud}^0$, given in Eq.~\ref{CKMel}, are a good approximation to the magnitude of the physical CKM angles (i.e., up to corrections of ${\cal O}(\delta^2)$, where $\delta$ is any one of the CP-violating phases).

\section{A physical framework for T-violation}

In the previous two sections we have described the
general features of the hidden symmetry and the
generic mechanism of breaking T-invariance and generating the corresponding light-quark 
masses in coincidence with the breaking of the hidden symmetry in the case of SM4.
In this section we would like to give a concrete physical example
(i.e., compatible with all relevant known data)
which is relatively simple analytically, therefore,
providing insight for the physical picture. Our chosen
setup below illustrates the
power of this mechanism in predicting the new mixing angles and phases associated with the 4th generation of quarks and the size of CP-violation of the theory.

As in the previous section, here also 
we consider a specific orientation for the hidden symmetry, where the direction of HS2 is partly fixed by setting $\zeta = \pi/2$ in each sector. The hidden symmetry is then broken by inserting the phases in the $12$ and $34$ elements of the mass matrix $M_0$, such that:

\begin{eqnarray}
\Delta M_z = (\Delta M)_{12} ~,~ \Delta M_t = (\Delta M)_{34} ~,
\end{eqnarray}
where $(\Delta M)_{ij}$ is defined in Eq.~\ref{deltaM}. Note that
with $\omega=\zeta=\pi/2$ we have $B_1 = s_\theta$, $B_2 = -c_\theta c_\phi$, $B_3=0$, $B_4 = -c_\theta s_\phi$, $C_1=0$, $C_2 = s_\eta s_\phi$, 
$C_3 = - c_\eta$ and $C_4 = - s_\eta c_\phi$ (see Eqs.~\ref{bi} and \ref{ci}). Thus,  the 
overall T-violating term, $\Delta M = \Delta M_z + \Delta M_t$, is given by:

\begin{eqnarray}
\Delta M =\left( {\tiny{ \begin{array}{cccc}
 0 & - m_{P_2} s_\theta c_\theta c_\phi \left( e^{i \delta_{12}} - 1 \right) & 0 & 0 \\
 - m_{P_2} s_\theta c_\theta c_\phi \left( e^{-i \delta_{12}} - 1 \right) & 0 & 0 & 0  \\
0 & 0 &  0 & m_{P_3} s_\eta c_\eta c_\phi \left( e^{i \delta_{34}} - 1 \right) \\
0 & 0 & m_{P_3} s_\eta c_\eta c_\phi \left( e^{-i \delta_{34}} - 1 \right) & 0 \end{array} }} \right) \label{deltaMzeta}~.
\end{eqnarray}

For simplicity and without loss of generality, we will further take
$s_{\phi} \ll 1$ for $\phi =\phi_d \sim \phi_u$ (recall that
$\cos(\phi_u - \phi_d) \sim V_{ud} \sim 1$ implying $\phi_u \sim \phi_d$, see previous section), which allows us to obtain a relatively compact analytical picture.
In particular, one simplification that arises with this choice, is that the perturbation in the $Q_1-P_1$ subspace, $\Delta m(Q_1,P_1)$ in Eq.~\ref{deltaQP}, is approximately diagonal so that
$m_- \approx \Delta m_{QQ}$, $m_+ \approx \Delta m_{PP}$ and the corresponding states are 
$v_-  \approx Q_1$, $v_+ \approx P_1$ in each sector. In particular, $\Delta M$ in Eq.~\ref{deltaMzeta}  generates the following light-quark masses (we now add the superscripts $d$ and $u$ to distinguish between the angles in the down and up-quark sectors):
\begin{eqnarray}
m_d &\approx& 2 m_b s_{\theta_d}^2 c_{\theta_d}^2 \left( 1 - \cos\delta_{12}^d \right) \label{md}~,\\
m_s &\approx& 2 m_{b^\prime} s_{\eta_d}^2 c_{\eta_d}^2 \left( 1 - \cos\delta_{34}^d \right) \label{ms}~,\\
m_u &\approx& 2 m_t s_{\theta_u}^2 c_{\theta_u}^2 \left( 1 - \cos\delta_{12}^u \right) \label{mu}~,\\
m_c &\approx& 2 m_{t^\prime} s_{\eta_u}^2 c_{\eta_u}^2 \left( 1 - \cos\delta_{34}^u \right) \label{mc}~.
\end{eqnarray}
where (see Eq.~\ref{zeromasses} and set $\zeta = \pi/2$):
\begin{eqnarray}
m_b \approx \beta_d,~m_{b^\prime} \approx \alpha_d,~m_t \approx \beta_u,
~m_{t^\prime} \approx \alpha_u ~.
\end{eqnarray}

As expected, we cannot reproduce the physical light-quark mass spectrum if any of the phases $\delta_{ij}$ above vanishes.
Note also that, since $\eta_u \sim \eta_d$ and
$\theta_u \sim \theta_d$ (see Eqs.~\ref{ctetd} and \ref{ceta}), we can also use the expressions in Eqs.~\ref{md}-\ref{mc} for the
light-quark mass terms to relate the phases in one sector to the phases in the other sector:
\begin{eqnarray}
\frac{\delta_{12}^d}{\delta_{12}^u} &\sim & \sqrt{\frac{m_d m_t}{m_u m_b}} \sim 10 ~,\\
\frac{\delta_{34}^d}{\delta_{34}^u} &\sim & \sqrt{\frac{m_s m_{t^\prime}}{m_c m_{b^\prime}}} \sim 0.3~,
\end{eqnarray}
where we have taken $m_{t^\prime}/m_{b^\prime} \sim 1$.

Finally, for our chosen orientation with $\zeta=\pi/2$ and $\phi << 1$, the $P_{q q^\prime}$ elements required to calculate
the imaginary terms of the $4 \times 4$ CKM elements (see
Eqs.~\ref{pijdef}-\ref{Vudfull}) are given by (to first order in $\delta_{ij}$):
\begin{eqnarray}
P_{d b} &=& m_b s_{\theta_d} c_{\theta_d} \sin\delta_{12}^d ~, \nonumber \\
P_{s b^\prime} &=& m_{b^\prime} s_{\eta_d} c_{\eta_d} \sin\delta_{34}^d  ~, \nonumber\\
P_{u t} &=& m_t s_{\theta_u} c_{\theta_u} \sin\delta_{12}^u  ~, \nonumber\\
P_{c t^\prime} &=& m_{t^\prime} s_{\eta_u} c_{\eta_u} \sin\delta_{34}^u
\label{pij} ~,
\end{eqnarray}
and all other $P_{q q^\prime}$ elements vanish.
Using the expressions for the light-quark masses in Eqs.~\ref{md}-\ref{mc}, we can re-express the elements of the perturbation matrix
$P_{q q^\prime}$ in Eq.~\ref{pij} above in terms of the CP-phases and the light-quark
masses:
\begin{eqnarray}
P_{d b} &\approx& \sqrt{m_d m_b} \cos\left( \frac{1}{2} \delta_{12}^d \right) ~, \nonumber\\
P_{s b^\prime} &\approx& \sqrt{m_s m_{b^\prime}} \cos\left( \frac{1}{2} \delta_{34}^d \right) ~, \nonumber\\
P_{u t} &\approx& \sqrt{m_u m_t} \cos\left( \frac{1}{2} \delta_{12}^u \right) ~, \nonumber\\
P_{c t^\prime} &\approx& \sqrt{m_c m_{t^\prime}} \cos\left( \frac{1}{2} \delta_{34}^u \right) \label{pij2}~,
\end{eqnarray}

\section{CP-invariants with four generations}

As in the SM3, CP-violation in the SM4 can also be parameterized
using CP-invariants a la the Jarlskog invariant $J_{SM}$ of the SM3 \cite{jarlskog}.
Indeed, as was shown in \cite{jarlskog}, the invariant CP-violation measure in the four quark families
case can be expressed in terms of four ``copies" similar to $J_{SM}$ (out of which only three
are independent):
$J_{123},~J_{124},~J_{134}$ and $J_{234}$, where the indices indicate
the generation number, i.e., in this language one identifies
$J_{SM}$ with $J_{123}$ even though these two CP-invariants are not quite the same
as $J_{SM}$ is no longer a valid CP-quantity in the SM4.

A generic derivation of the four $J_{ijk}$ copies in terms of the
quark masses and CKM mixing angles is quite complicated and we are unable to give it
in a compact analytical format.
There are several
useful general formulations in the literature for the parametrization of CP-violation in the SM4
\cite{jarlskog,CPinvariants}, but none is at the level of simplification
required for an analytical study of CP-violation in our model.
A numerical calculation/study of the CP-violating quantities in our model is, however,
straight forward following the prescription of the previous sections.
This will be presented elsewhere \cite{our-2nd-FLpaper}.

On the other hand,
as was observed more than 10 years ago \cite{branco} and noted again
recently in \cite{CPbaryo1}, in the chiral limit $m_{u,d,s,c} \to 0$,
CP-violation
in the SM4 effectively ``shrinks" to
the CP-violation picture of a three generation model involving
the 4th generation heavy quarks.
This chiral limit, which is in the spirit of our current study,
is clearly applicable at high-energies
of the EW-scale and above. Moreover,
it allows us to derive a compact analytical estimate for
the expected size of CP-violation in our model.\footnote{Note that, although
their is no CP-violation in our model in the chiral limit $m_{u,d,s,c} \to 0$ (which is our zeroth-order approximation), we can use the CP-violating quantities obtained in \cite{branco} in this limit, since those are given in terms of the physical mixing angles. In our model, the imaginary parts of these mixing angles are proportional
to the very small light-quark masses.}

As was shown in \cite{branco}, in the chiral limit there is no CP-violation
within the three families SM3 and so all CP-violating effects are attributed to the
new physics - in our case, to the fourth generation of quarks.
The key CP-violating quantity in this limit can be written as \cite{branco}:
\begin{eqnarray}
 J_{SM4} = {\rm Im} \left( V_{tb} V_{t^\prime b}^\star V_{t^\prime b^\prime} V_{t b^\prime}^\star \right)
~,
 \label{b2} ~.
\end{eqnarray}
since this is the only CP-violating quantity that survives when one takes the
limit $m_{u,d,s,c} \to 0$.

Thus, in order to get some insight
for the expected size of CP-violation in our model, it is sufficient to
derive an estimate for $J_{SM4}$.
In particular, we will calculate $J_{SM4}$ for the specific orientation
used in the previous section, i.e., for the case $\zeta=\pi/2$ and
$\phi \ll 1$.

Using the $P_{q q^\prime}$ factors of Eq.~\ref{pij2} and based on
Eq.~\ref{Vudfull}, we can calculate (to ${\cal O}(\delta)$) the relevant
complex CKM elements which enter $J_{SM4}$ in Eq.~\ref{b2}:
\begin{eqnarray}
V_{tb} &\approx& V_{tb}^0 + i \left[ V_{td}^0 \sqrt{ \frac{m_d}{m_b} }
\cos\left(\frac{1}{2} \delta_{12}^d \right) -
V_{ub}^0 \sqrt{ \frac{m_u}{m_t} }
\cos\left(\frac{1}{2} \delta_{12}^u \right) \right] ~, \\
V_{t^\prime b} &\approx& V_{t^\prime b}^0 + i \left[ V_{t^\prime d}^0 \sqrt{ \frac{m_d}{m_b} }
\cos\left(\frac{1}{2} \delta_{12}^d \right) -
V_{cb}^0 \sqrt{ \frac{m_c}{m_{t^\prime}} }
\cos\left(\frac{1}{2} \delta_{34}^u \right) \right] ~, \\
V_{t b^\prime} &\approx& V_{t b^\prime}^0 + i \left[ V_{t s}^0 \sqrt{ \frac{m_s}{m_{b^\prime}} }
\cos\left(\frac{1}{2} \delta_{34}^d \right) -
V_{u b^\prime}^0 \sqrt{ \frac{m_u}{m_{t}} }
\cos\left(\frac{1}{2} \delta_{12}^u \right) \right] ~, \\
V_{t^\prime b^\prime} &\approx& V_{t^\prime b^\prime}^0 + i \left[ V_{t^\prime s}^0 \sqrt{ \frac{m_s}{m_{b^\prime}} }
\cos\left(\frac{1}{2} \delta_{34}^d \right) -
V_{c b^\prime}^0 \sqrt{ \frac{m_c}{m_{t^\prime}} }
\cos\left(\frac{1}{2} \delta_{34}^u \right) \right] ~.
\end{eqnarray}

We can now estimate the size of CP-violation in our model, which can emanate
in high-energy processes involving $t^\prime$ and $b^\prime$ exchanges. In particular,
since the zeroth-order CKM elements are a good approximation for the magnitude of physical elements, we set $V_{ij}^0 \sim V_{ij}$ and use the results and relations
obtained for the CKM elements in the previous sections
(see Eqs.~\ref{eqV1}-\ref{eqV6}):
$V_{tb} \sim 1$, $V_{t^\prime b^\prime} \sim V_{cs} \sim 1$,
$V_{t^\prime b} \sim V_{u b^\prime} \times (V_{cb}/V_{cd})$ and
$V_{t b^\prime} \sim V_{u b^\prime} \times (V_{ts}/V_{us})$.
We then obtain:
\begin{eqnarray}
J_{SM4} \approx
&& V_{u b^\prime} \frac{V_{ts}}{V_{us}} \times \left[
V_{c b}  \sqrt{ \frac{m_c}{m_{t^\prime}} }
\cos\left(\frac{1}{2} \delta_{34}^u \right) -
V_{u b^\prime} \sqrt{ \frac{m_d}{m_{b}} }
\cos\left(\frac{1}{2} \delta_{12}^d \right) \right] + \nonumber \\
&&V_{u b^\prime} \frac{V_{cb}}{V_{cd}} \times \left[
V_{u b^\prime}  \sqrt{ \frac{m_u}{m_{t}} }
\cos\left(\frac{1}{2} \delta_{12}^u \right) -
V_{ts} \sqrt{ \frac{m_s}{m_{b^\prime}} }
\cos\left(\frac{1}{2} \delta_{34}^d \right) \right]
\label{moreb2}~.
\end{eqnarray}
Setting $V_{cb} \sim - V_{ts} \sim A \lambda^2$ and 
$V_{ts}/V_{us} \sim V_{cb}/V_{cd} \sim - A \lambda$
and (consistent with their measured values \cite{PDG}, where $A \sim 0.81$ and $\lambda = 0.2257$ is the Wolfenstein parameter),
and taking $V_{u b^\prime} \sim V_{cb} \sim A \lambda^2$
and $m_{t^\prime} \sim 2 m_t$, $m_{b^\prime} \sim m_{t^\prime} - 55~{\rm GeV}$,
consistent with the electroweak precision tests \cite{kribs,0902.4883},
we obtain:
\begin{eqnarray}
\left| J_{SM4} \right| \sim A^3 \lambda^5 \times
\left[ 
\sqrt{ \frac{m_u}{m_{t}} }
+
\sqrt{ \frac{m_c}{m_{t^\prime}} } 
-
\sqrt{ \frac{m_d}{m_{b}} }
+
\sqrt{ \frac{m_s}{m_{b^\prime}} }
\right] \sim
10^{-5} \label{jsm4} ~,
\end{eqnarray}
where we have used $\cos(\delta_{12}^d/2) \sim \cos(\delta_{34}^d/2) \sim \cos(\delta_{12}^u/2) \sim \cos(\delta_{34}^u/2) \sim 1$ for the numerical estimate (see below). Indeed, with the above chosen values for
the CKM elements and the 4th generation quark masses, all the four
phases are fixed by the requirement that they reproduce the corresponding
light-quark masses as given in Eqs.~\ref{md}-\ref{mc}. In particular, according to Eqs.\ref{md}-\ref{mc} and the relations between the hidden symmetry angles and the CKM elements as given by Eqs.~\ref{eqV1}-\ref{eqV4}, we have:
\begin{eqnarray}
\cos\left(\delta_{23}^u \right) &\sim& 1 - \frac{m_c}{2 m_{t^\prime} \frac{V_{u b^\prime}^2}{V_{cd}^2}} \sim 0.945  ~, \\
\cos\left(\delta_{12}^d \right) &\sim& 1 -\frac{m_d}{2 m_{b}
\frac{V_{c b}^2}{V_{cd}^2} }  \sim 0.98 ~, \\
\cos\left(\delta_{23}^d \right) &\sim& 1 -\frac{m_s}{2 m_{b^\prime}
\frac{V_{u b^\prime}^2}{V_{us}^2} } \sim 0.995 ~, \\
\cos\left(\delta_{12}^u \right) &\sim& 1 -\frac{m_u}{2 m_{t}
\frac{V_{ts}^2}{V_{us}^2} } \sim 0.9998 ~,
\end{eqnarray}
consistent with our perturbative description of CP-violation.

From Eq.~\ref{jsm4} we see that
as the CP-violating phases $\delta_{12}^d,\delta_{34}^u \to 0$,
both $m_d$ and $m_c$ approach zero and, therefore, also $J_{SM4} \to 0$.
Note also that, for our chosen orientation of the hidden symmetry, we have
$J_{SM4} \sim 10^{-5} \sim J_{SM}$, i.e.,
the SM4 analogue of the SM3's Jarlskog invariant at high-energies
and the original measured SM3's Jarlskog invariant are of similar size.
These results demonstrate the highly predictive power of our model
for the description of CP-violation and the generation of the light-quark masses in the SM4. In particular, once the magnitude
of the mixing angles and the masses of the 4th generation quarks are measured,
our model gives a very distinct prediction for the expected size of CP-violation
in the SM4, which can be directly confirmed at high-energy collider experiments.
In a forthcoming paper \cite{our-2nd-FLpaper}, we will perform a full numerical study and scan the complete range of the free parameter space of our model, subject to
the relevant existing data. We will also suggest ways to test our model in the upcoming
LHC and the future machines such as a Super-B factory and the International
Linear Collider.

\section{Summary}

Motivated by the recent hints of CP anomalies in the B-system and by
the idea of Friedberg and Lee (FL) in \cite{FL}, we have
presented a new framework for CP-violation and the generation of the light-quark masses in the SM with four families - the SM4.

We have applied the basic ingredients of the FL mechanism to the SM4 case,
by constructing an extended (double) hidden symmetry suitable for four
families which defines the zeroth-order states in the up and down-quarks sectors and
which ensures T-invariance.
We then outlined the breaking mechanism of both the hidden symmetry and
T-invariance in the SM4 case, from which we obtained the CP-violating measure
and the physical states in this model.
We have shown that this mechanism, when applied to the SM4, can be highly predictive
and can be tested in future experiments. In particular, we gave one physically relevant  example for
the predictive power of our model
by choosing a specific orientation of the hidden symmetry. This allowed us to
analytically derive the physical (observed) quark states, and to give a prediction
for the size of the mixing angles between the 4th generation and the 1st three generations of the SM3 and for the size of CP-violation associated with the 4th generation quarks.

A complete numerical study of our model,
which explores the full phase-space of viable hidden symmetries for the SM4
and the corresponding range of the expected size of CP-violation and
of the 4th generation mixing angles, is in preparation and will
be presented in \cite{our-2nd-FLpaper}.

\bigskip

{\it \bf Acknowledgments:}
We thank Gad Eilam for discussions. The work of AS is supported
in part by the US DOE contract No. DE-AC02-98CH10886.

\end{document}